\pdfoutput=1
\documentclass[conference]{IEEEtran}

\usepackage{cite}
\usepackage{amsmath,amssymb,amsfonts}
\usepackage{algorithmic}
\usepackage{graphicx}
\usepackage{textcomp}
\usepackage{xcolor}
\usepackage{enumitem}
\usepackage[hidelinks]{hyperref}

\def\BibTeX{{\rm B\kern-.05em{\sc i\kern-.025em b}\kern-.08em
    T\kern-.1667em\lower.7ex\hbox{E}\kern-.125emX}}

\begin{document}

\title{Scalable Multi-GPU Simulation of 3D Multicellular Growth
with RNN-Based Workload Balancing}

\author{
\IEEEauthorblockN{
Matvey Moisseyev\IEEEauthorrefmark{1}\IEEEauthorrefmark{2},
Huijing Du\IEEEauthorrefmark{2},
Dandan Zheng\IEEEauthorrefmark{3},
Chi Zhang\IEEEauthorrefmark{4},
and Hongfeng Yu\IEEEauthorrefmark{1}\IEEEauthorrefmark{6}
}

\IEEEauthorblockA{
\IEEEauthorrefmark{1}School of Computing, University of Nebraska--Lincoln\\
\IEEEauthorrefmark{2}Department of Mathematics, University of Nebraska--Lincoln\\
\IEEEauthorrefmark{3}Department of Radiation Oncology, University of Rochester Medical Center\\
\IEEEauthorrefmark{4}School of Biological Sciences, University of Nebraska--Lincoln\\
\IEEEauthorrefmark{6}Holland Computing Center, University of Nebraska--Lincoln
}
}

\maketitle

\begin{abstract}
Detailed multicellular growth simulations based on subcellular element models (SEMs) can capture complex tissue development, but their element-level interactions impose substantial computational cost. 
This work presents a scalable multi-GPU framework for three-dimensional multicellular growth simulation that combines GPU acceleration, spatial binning, domain decomposition, and workload-aware partitioning. 
A central challenge is that cell movement, growth, and division continuously reshape the spatial workload distribution, causing initially balanced partitions to become inefficient over time. 
To address this problem, we introduce an RNN-based load-balancing controller that observes recent per-rank execution times and partition states and learns residual corrections to a reactive boundary-adjustment rule. 
The controller is trained offline in a differentiable surrogate of the load-balancing loop with randomized workload dynamics, requiring no measured execution traces for training. 
We evaluate the framework in terms of single-GPU acceleration, multi-GPU computation scaling, controller-level load-balancing behavior, and end-to-end simulation performance, with comparisons against static partitioning, reactive load balancing, and conventional time-series prediction baselines. 
A representative embryonic epidermal development use case further demonstrates the type of spatially and temporally evolving workload targeted by the framework. 
In our evaluation, GPU acceleration with spatial binning accelerates the interaction computation by roughly three orders of magnitude over a serial CPU baseline. 
RNN-guided load balancing reduces the mean global imbalance from \(11.3\%\) under static partitioning to \(3.5\%\), lowers end-to-end runtime by \(9.0\%\) relative to static partitioning, and reduces slice migration by \(7.7\times\) compared with the reactive baseline, showing that history-aware control can improve workload balance while avoiding unnecessary repartitioning.
\end{abstract}

\begin{IEEEkeywords}
multicellular simulation, subcellular element model, multi-GPU computing, dynamic load balancing, domain decomposition, high-performance computing
\end{IEEEkeywords}

\section{Introduction}
\label{sec:introduction}

Multicellular tissue development emerges from interactions among individual cells, including growth, division, migration, adhesion, and cell-fate transitions. 
Computational models provide a complementary approach to experiments by making these mechanisms explicit and controllable, enabling researchers to investigate how local cell behaviors give rise to tissue-scale structure over time. 
In particular, three-dimensional cell-based models can directly represent individual cells and their mechanical interactions, making them well suited for studying complex developmental processes~\cite{osborne2017comparing,fletcher2017mechanocellular}.

Greater biological and mechanical detail, however, comes with substantial computational cost. 
Subcellular element models (SEMs) provide a cell-resolved formulation in which each cell is discretized into multiple interacting computational elements~\cite{newman2005modeling}. 
This representation captures local mechanical interactions at a finer scale than center-based cell models, but it also increases the number of force evaluations required during each simulation step. 
As cells grow, divide, and move, both the number and the spatial distribution of subcellular elements change over time. 
Consequently, the cost of local force evaluation varies across the tissue and across the simulation trajectory, creating a computational workload that is both large and dynamically evolving.

The locality and inherent parallelism of element interactions make GPUs well suited for accelerating these computations. 
Earlier work demonstrated substantial GPU acceleration for three-dimensional epidermal SEM simulations~\cite{christley2010gpu}, and other GPU-based frameworks have enabled increasingly large multicellular and agent-based simulations~\cite{germann2019yalla,richmond2023flame}. 
A single GPU, however, remains constrained by its computational capacity and device memory. 
A multi-GPU implementation of an evolving tissue model therefore requires spatial domain decomposition and communication of boundary data between neighboring ranks, while preserving the model's local interactions during synchronization.

An additional challenge is that the workload is neither spatially uniform nor temporally stationary. 
Cell growth, division, and movement continuously change local element density and therefore the number of interaction calculations required in different regions of the domain. 
As a result, a decomposition that is well balanced at one stage of the simulation may become imbalanced later. 
Because ranks synchronize during execution, overall performance is constrained by the slowest rank. 
Reactive load balancing can adjust partition boundaries after imbalance is observed, but it does not explicitly exploit the temporal evolution of the workload. 
This motivates a history-aware strategy that uses recent workload trends to guide future boundary adjustments.

In this work, we develop a scalable multi-GPU framework for three-dimensional multicellular growth simulation based on an epidermal SEM. 
The biological and mathematical model builds on prior multiscale epidermal development work that integrates SEM-based mechanics with lineage progression, adhesion rules, and extracellular signaling during tissue formation~\cite{du2018multiscale}. 
Our framework combines GPU-parallel element interaction calculations with spatial binning, partitions the simulation domain across multiple GPUs using ghost regions for boundary interactions, and uses an explicitly computed workload score to construct the initial decomposition. 
To accommodate changes in workload as the tissue evolves, we further introduce an RNN-guided load-balancing controller that observes recent per-rank load and partition-state history at each partition boundary and learns residual corrections to a reactive boundary-adjustment rule.

The main contributions of this work are:

\begin{itemize}[leftmargin=*]
    \item We develop a multi-GPU framework for three-dimensional multicellular growth simulation based on a subcellular element model, combining GPU-parallel force evaluation, spatial binning, and distributed spatial domain decomposition.

    \item We introduce a workload-aware initial partitioning method that computes interaction workload from cell element counts and local element density to distribute the simulation workload more evenly across GPU ranks.

    \item We develop an RNN-guided load-balancing controller that learns residual corrections to a reactive boundary-adjustment rule using recent workload and partition-state history.
    
    \item We evaluate the framework using single-GPU, multi-GPU, controller-level, and end-to-end simulation experiments.
\end{itemize}

\section{Related Work}
\label{sec:related}

\subsection{Cell-Based and Subcellular-Element Modeling}

Tissue growth can be simulated either with continuum models or with cell-resolved models. In this work, we focus on cell-resolved approaches, where individual cells maintain their own states and interact with nearby cells through local rules for proliferation, migration, adhesion, and mechanical contact. This representation is particularly useful for epidermal development and wound-healing simulations, where tissue structure is shaped by many local cell-level interactions~\cite{du2018multiscale,wang2019woundhealing}. Different formulations make different assumptions about cell geometry and neighborhood structure, including lattice-based, center-based, vertex-based, and off-lattice approaches~\cite{osborne2017comparing,fletcher2022seven}. Frameworks such as CellSys and PhysiCell provide general-purpose environments for three-dimensional cell-based simulation~\cite{hoehme2010cellsys,ghaffarizadeh2018physicell}.

In the SEM formulation, each cell is discretized into multiple computational elements whose local interactions determine the cell's effective geometry and mechanical response~\cite{newman2005modeling,sandersius2008rheology}. This cell-resolved representation has been used to study how element-level mechanics and active cellular processes give rise to collective multicellular behavior~\cite{sandersius2011emergent,newman2008gridfree,revell2019sorting}. For large tissue simulations, however, the same modeling detail introduces substantial computational cost because each cell update requires many short-range element--element interaction evaluations.

SEM-based formulations have also been applied to epidermal development. Gord et al. used an anisotropic subcellular-element model to study epidermal stratification and asymmetric cell division~\cite{gord2014epidermal}. Du et al. subsequently developed a multiscale epidermal model that combines subcellular mechanics with cell-lineage progression, selective adhesion, and extracellular signaling~\cite{du2018multiscale}. The present work builds on this model. Its element-level resolution provides detailed mechanical behavior, but also creates a substantial computational workload because each biological cell contains multiple interacting elements and the element distribution changes continuously during tissue growth.

\subsection{Parallel and GPU-Accelerated Multicellular Simulation}

Parallel computing has also been explored as a way to increase the size and complexity of multicellular simulations. Some systems provide general parallel infrastructure for cell-based modeling, such as BioCellion, while others, such as PhysiCell, emphasize efficient simulation of large three-dimensional cell populations~\cite{kang2014biocellion,ghaffarizadeh2018physicell}. Additional efforts have targeted distributed execution for more specific biological processes, including cell--cell communication and coupled models of cell physics and signaling~\cite{coulier2018orchestral,merchant2023dense}. These studies demonstrate the value of parallel execution for biological simulation. Our work builds on this direction in the context of dynamically growing SEM tissues. 
In this setting, local subcellular interactions require communication across GPU boundaries, while cell growth and division continually reshape the spatial distribution of computation. As a result, maintaining balanced workloads across GPUs becomes a central challenge.

GPUs are particularly attractive for multicellular mechanics because many local interaction calculations can be evaluated concurrently. Christley et al. developed GPU algorithms for a three-dimensional epidermal SEM and demonstrated substantial acceleration of element-level computation~\cite{christley2010gpu}. GPU acceleration has also been explored for off-lattice multicellular models. For example, ya$\parallel$a demonstrated that force-based multicellular morphogenesis simulations can be mapped effectively to GPU architectures~\cite{germann2019yalla}. FLAME GPU 2 provides a general GPU-based framework for large agent-based simulations~\cite{richmond2023flame}, and Borau et al. applied FLAME GPU 2 to cell--microenvironment simulation with cell--cell and cell--matrix interactions~\cite{borau2024flamegpu2}. These GPU-based studies motivate the use of accelerator architectures for multicellular simulation, while the present work focuses on distributing a dynamically growing SEM tissue across multiple GPUs with runtime load balancing.

\subsection{Dynamic Load Balancing and Workload Prediction}

Dynamic load balancing is important in spatial and particle-based simulations because the amount of computation assigned to each partition can change substantially during execution. Begau and Sutmann developed adaptive irregular domain decomposition for particle simulations with strongly nonuniform and dynamically changing particle distributions~\cite{begau2015adaptive}. In distributed multi-GPU particle tracking, Yang et al. showed that evolving particle distributions can reduce parallel efficiency and used dynamic load balancing to redistribute work across processing elements~\cite{yang2022waterage}. Runtime load balancing has been explored in other particle-based simulations, including CPU--GPU CFD--DEM and multi-GPU smoothed particle hydrodynamics~\cite{zhu2023dynamic,ahn2026dynamic}. These studies show that adapting domain boundaries can improve parallel efficiency, but SEM-based tissue growth introduces different sources of imbalance, including cell division, local density changes, and short-range element interactions.

Predictive load balancing aims to adjust the partitioning before workload imbalance becomes severe. This idea has been studied in several simulation contexts, including molecular simulation and HLA-based distributed simulation, where predicted workload information is used to guide partition adjustment or object migration~\cite{guzman2017predictive,degrande2011predictive}. Other studies have used time-series models to forecast workload changes and support migration decisions in distributed simulations~\cite{alkharboush2013holt,degrande2017timeseries}. Recent cloud-computing studies further show that recurrent and attention-based models can capture nonlinear temporal load patterns~\cite{predic2024cloud}. Although these methods demonstrate the usefulness of prediction for load balancing, they address workload structures that differ from spatial scientific simulations.

In SEM-based tissue growth, workload evolves with cell growth, cell division, and changes in local element density. These changes create spatially localized imbalance that affects both GPU computation and inter-GPU boundary exchange. In this work, we investigate an RNN-guided dynamic load-balancing strategy that uses recent workload history and partition-state information to adapt the spatial decomposition during multi-GPU multicellular growth simulation.

\section{Framework Overview and Multi-GPU Simulation}
\label{sec:framework}

\begin{figure}[t]
    \centering
    \includegraphics[width=1.0\linewidth]{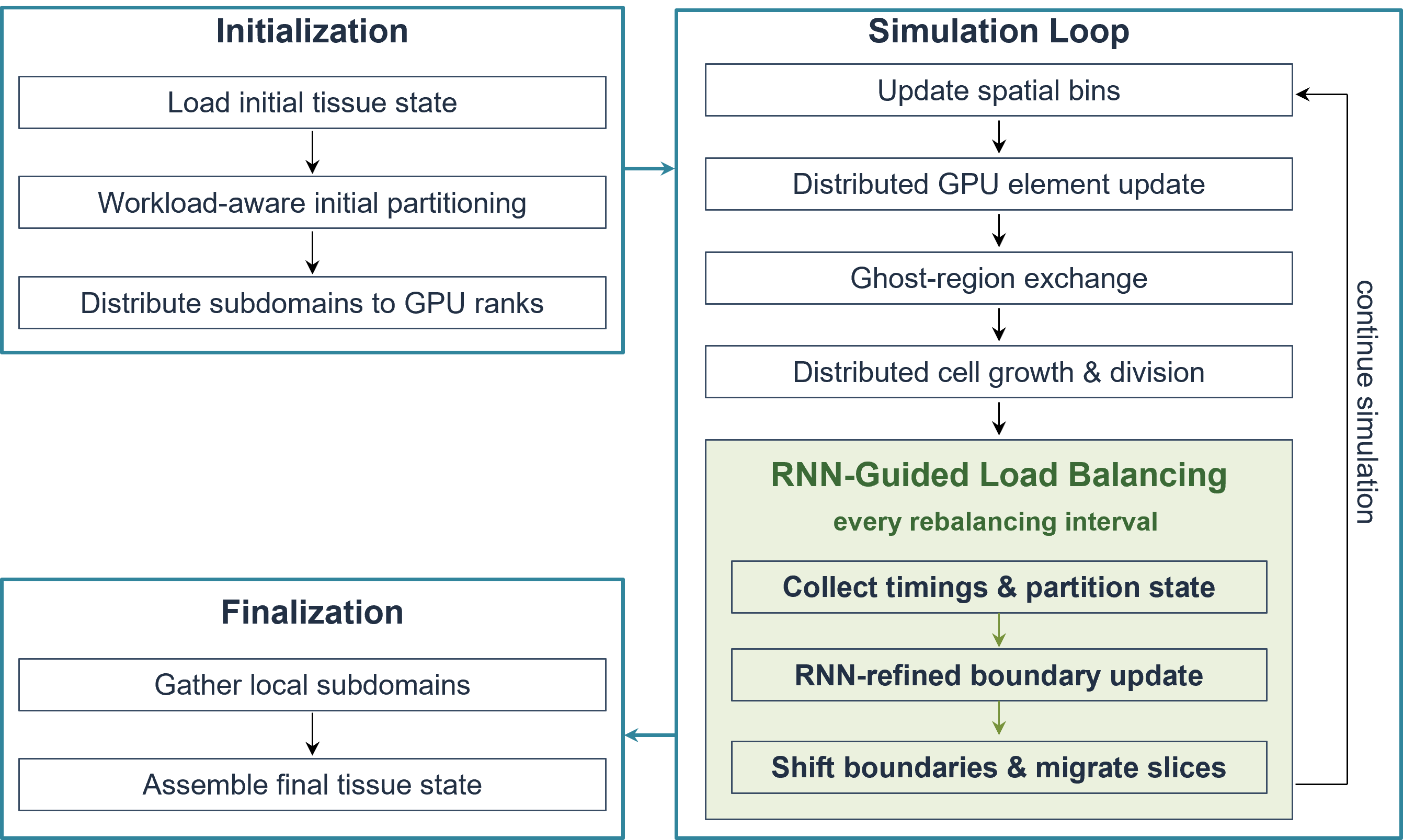}
    \caption{Overview of the proposed multi-GPU multicellular growth simulation framework. The framework initializes the simulation with workload-aware partitioning, distributes subdomains across GPU ranks, and iteratively performs spatial binning, distributed GPU element updates, ghost-region exchange, and cell growth and division. At each rebalancing interval, timing and partition-state information are used by the RNN-guided controller to update partition boundaries and migrate slices. 
The final tissue state is assembled from the local subdomains.}
    \label{fig:framework}
\end{figure}

The proposed framework builds on a previously developed multiscale model of epidermal tissue formation~\cite{du2018multiscale}. 
Figure~\ref{fig:framework} provides an overview of the full workflow, including workload-aware initialization, GPU-parallel element updates, ghost-region exchange, and RNN-guided runtime repartitioning. 
This section first summarizes the epidermal SEM components that determine the computational workload, and then describes the multi-GPU simulation framework used to accelerate element-level interactions and distribute the spatial domain across GPU ranks. 
The RNN-guided dynamic load-balancing method is described separately in Section~\ref{sec:loadbalance}.

\subsection{Epidermal Subcellular Element Model}

Following the epidermal SEM used in prior work~\cite{du2018multiscale}, each biological cell is represented by multiple interacting subcellular elements. 
For an element $\alpha_i$ in cell $\alpha$, its position $Y_{\alpha_i}$ is updated according to forces from nearby elements and external structures:
\begin{equation}
\frac{dY_{\alpha_i}}{dt}
=
-\nabla_{\alpha_i}
\sum_{\beta_j \in \mathcal{I}}
V_{\mathrm{ele}}
\left(
\left|Y_{\alpha_i}-Y_{\beta_j}\right|
\right)
-
\nabla_{\alpha_i}
V_{\mathrm{ext}}(Y_{\alpha_i}),
\end{equation}
where $\mathcal{I}$ is the set of neighboring elements considered in the force evaluation, $V_{\mathrm{ele}}$ includes intra-cellular elastic and inter-cellular adhesive--repulsive interactions, and $V_{\mathrm{ext}}$ models external adhesion such as basement-membrane effects. 
All element--element interactions are evaluated within a prescribed cutoff radius, which defines both the local GPU neighborhood search and the ghost-region width for inter-GPU communication.

The model also supports cell growth and division. 
Cell growth increases the number of subcellular elements assigned to a cell, while division partitions the elements into two daughter cells according to the division rules of the epidermal model~\cite{du2018multiscale}. 
As cells grow, divide, and move, both the number and spatial distribution of elements change over time, creating the dynamic workload addressed in this work.

\subsection{Computational Characteristics}

The dominant computational cost arises from element--element force evaluation. A naive all-pairs implementation requires $O(n^2)$ candidate interaction tests for $n$ elements. Because the interaction potentials are truncated beyond a finite cutoff distance, however, only spatially nearby elements can contribute to the force acting on a given element.

The workload is also spatially nonuniform and evolves throughout the simulation. Dense regions require more interaction evaluations than sparse regions, while cell movement, growth, and division continually alter the number and spatial distribution of elements. Consequently, a spatial decomposition that is initially balanced may become inefficient as the tissue evolves.

These characteristics create two related computational requirements: efficient evaluation of local element interactions and scalable distribution of a dynamically changing spatial workload across multiple GPUs.

\subsection{GPU Parallelization and Spatial Binning}

Element movement is accelerated on the GPU. To avoid the quadratic cost of all-pairs candidate comparisons, the three-dimensional domain is organized into a regular grid of cubic bins whose size is determined by the interaction cutoff distance.

Each element examines interaction candidates only in its own bin and the 26 neighboring bins. The resulting $3\times3\times3$ neighborhood search avoids comparisons with the entire element set. With this spatial-binning strategy, under bounded local density, the number of candidate interactions per element remains approximately constant, allowing the interaction calculation to scale linearly with the number of elements rather than quadratically.

\subsection{Multi-GPU Domain Decomposition}

To extend the simulation beyond a single GPU, the spatial domain is divided into contiguous slices along one spatial dimension, with groups of slices assigned to individual GPU ranks. The partitioning dimension can be selected according to the geometry of the simulated tissue.

Interactions near partition boundaries require information from neighboring ranks. Each rank therefore maintains a ghost region containing elements within the interaction cutoff distance of adjacent partitions. Because interactions are truncated at the cutoff distance, boundary information is exchanged only between neighboring ranks rather than globally across all GPUs.

This decomposition preserves local element interactions while distributing both computational workload and element data across multiple devices.

\subsection{Workload-Aware Initial Partitioning}
\label{sec:initial_partition}

A uniform geometric decomposition does not necessarily balance the computational workload, because subdomains with the same geometric size may contain different numbers of cells and different local interaction densities. In our subcellular element model, the cost of updating a cell depends on both the number of subcellular elements in the cell and the number of neighboring elements within the interaction cutoff. We therefore compute a workload score before constructing the initial partition.

For each cell $i$, the workload score is
\begin{equation}
w_i = n_i n_{\mathcal{N}(i)},
\end{equation}
where $n_i$ is the number of subcellular elements belonging to cell $i$, and $n_{\mathcal{N}(i)}$ is the number of elements contained in the surrounding $3\times3\times3$ bin neighborhood.

\begin{figure}[t]
    \centering
    \includegraphics[width=0.80\columnwidth]{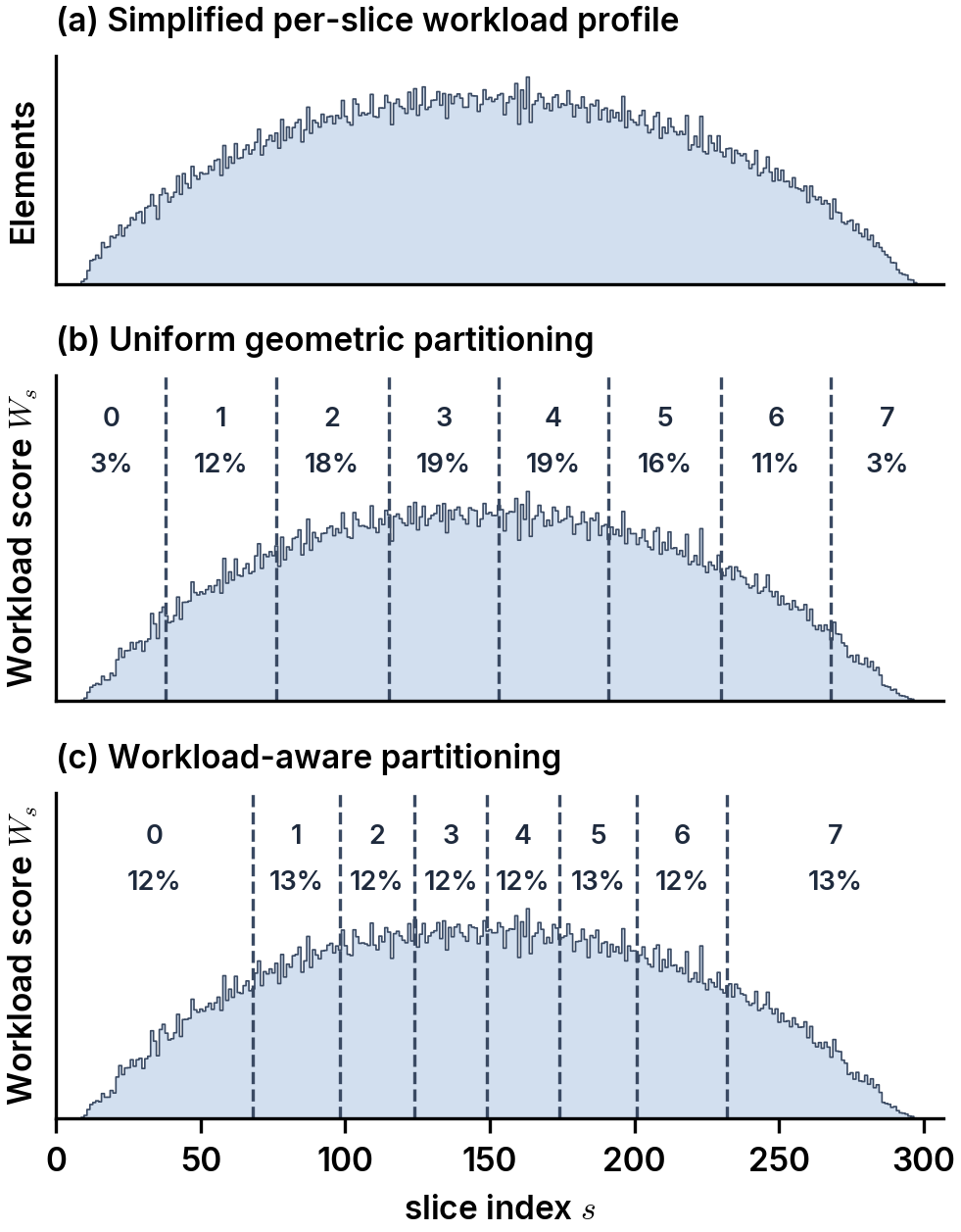}
    \caption{Schematic illustration of workload-aware initial partitioning. 
    (a) The number of subcellular elements varies across slices along the partitioning axis, producing a spatially nonuniform workload. 
    (b) Uniform geometric partitioning assigns equal-width slice ranges to GPU ranks, which can produce unequal workload shares when the workload is spatially nonuniform. 
    (c) The workload-aware method selects partition boundaries using the accumulated workload score \(W_s\) so that each rank receives approximately the same total workload. 
    This example uses 8 GPU ranks, labeled 0--7. 
    Percentages are rounded workload shares assigned to each GPU rank. 
    Practical workload profiles are often more irregular.}
    \label{fig:initial_partition}
\end{figure}

The simulation domain is divided along the partitioning axis into a sequence of thin slices. Each slice has a thickness determined by dividing the spatial domain with padding into a set number of slices. Then the workload of slice $s$ is
\begin{equation}
W_s = \sum_{i \in s} w_i.
\end{equation}
The sequence
\begin{equation}
\mathbf{W}
=
[W_1,W_2,\ldots,W_m]
\end{equation}
forms a one-dimensional workload histogram with $m$ slices. Figure~\ref{fig:initial_partition} illustrates the workload-aware partitioning process.

Contiguous slices are grouped so that the accumulated workload of each GPU rank is approximately equal,
\begin{equation}
W_r = \sum_{s \in P_r} W_s,
\end{equation}
where $P_r$ is the set of slices assigned to rank $r$.

This workload-aware strategy provides a more balanced initial decomposition than uniform geometric partitioning when the computational workload is spatially heterogeneous. 
However, this initial partition only reflects the workload distribution at the beginning of the simulation. 
As the tissue evolves, runtime imbalance can still emerge, motivating the dynamic load-balancing method described in Section~\ref{sec:loadbalance}.

\section{Predictive Dynamic Load Balancing}
\label{sec:loadbalance}

\subsection{Motivation and Design Rationale}

A conventional reactive strategy can correct imbalance after it is observed from GPU execution times, but such adjustments respond only after performance has already degraded.
Moreover, workload evolution is temporally correlated: recent workload and partition states can provide information about how local imbalance may change in subsequent simulation intervals.

\begin{figure}[t]
    \centering
    \includegraphics[width=0.98\columnwidth]{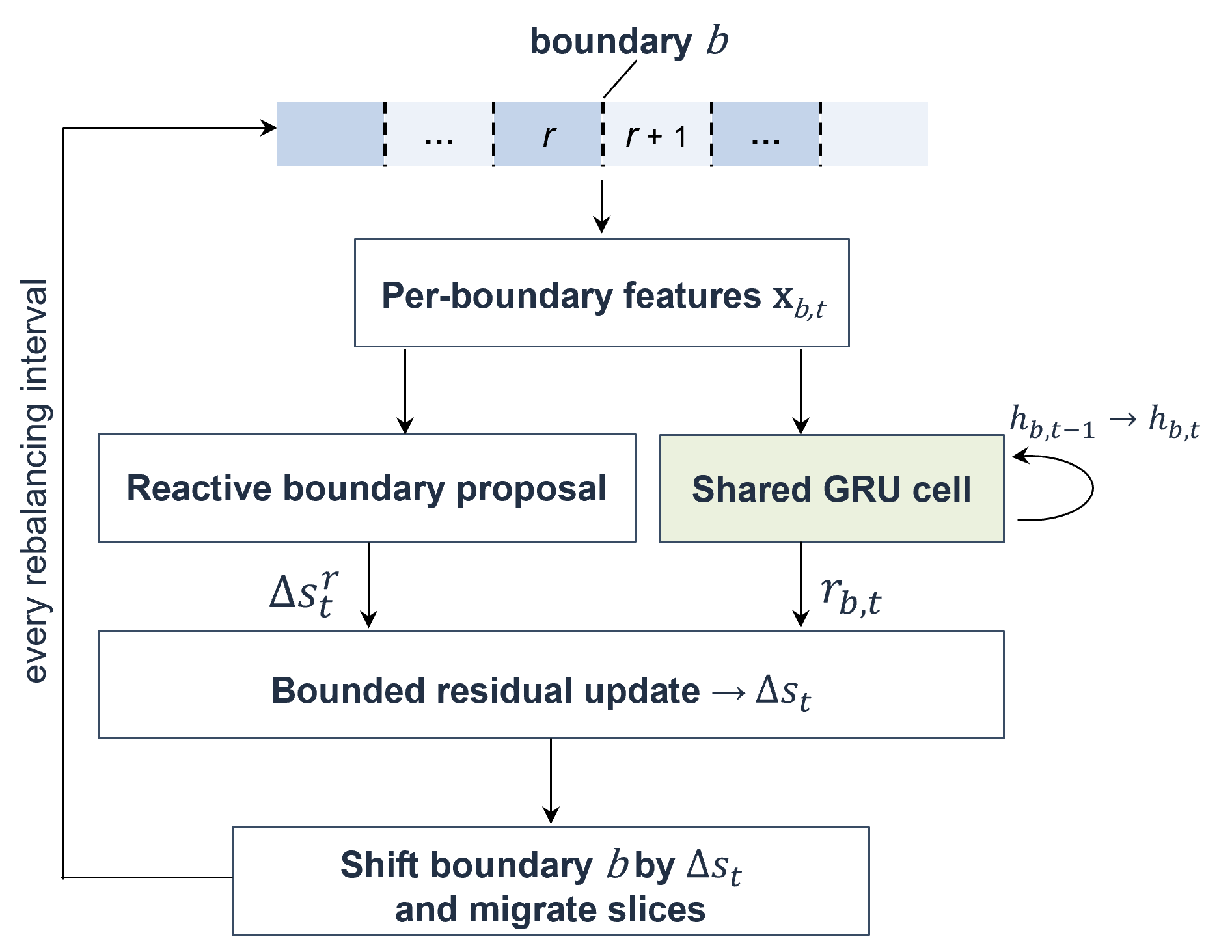}
    \caption{RNN-guided dynamic load balancing. 
    For each interior partition boundary, timing, workload-history, and partition-state features are collected and passed to both a reactive boundary proposal and a shared recurrent controller. 
    The RNN outputs a residual correction that is combined with the reactive proposal to produce a bounded boundary adjustment. 
    The corrected adjustment shifts the partition boundary and migrates slices between neighboring GPU ranks.}
    \label{fig:rnn_rebalancing}
\end{figure}

Our approach therefore uses recent workload history and partition-state information to guide each boundary adjustment. 
A recurrent neural network (RNN) maintains a hidden state for each partition boundary and outputs a residual correction to the reactive boundary proposal. 
The corrected adjustment shifts the boundary and migrates slices between neighboring GPU ranks. 
Figure~\ref{fig:rnn_rebalancing} summarizes this rebalancing workflow.

\subsection{Reactive Baseline}

The baseline runtime method monitors the computation times of neighboring ranks. 
For two ranks adjacent to a partition boundary, we define the local imbalance at rebalancing step $t$ as
\begin{equation}
I_t =
\frac{T_t^{+}-T_t^{-}}
{\left(T_t^{+}+T_t^{-}\right)/2},
\end{equation}
where $T_t^{-}$ and $T_t^{+}$ are the measured computation times of the two ranks adjacent to the boundary.

A boundary adjustment is triggered when
\begin{equation}
|I_t| > \tau,
\end{equation}
where $\tau$ is the imbalance threshold.

When rebalancing is triggered, the partition boundary is shifted to transfer workload from the slower rank to the faster rank. 
The adjustment magnitude is
\begin{equation}
|\Delta s_t^{\mathrm{r}}|
=
\alpha N_{\mathrm{slow}}
\frac{|T_t^{+}-T_t^{-}|}{2T_t^{\mathrm{slow}}},
\end{equation}
where \(\Delta s_t^{\mathrm{r}}\) is the reactive boundary proposal, 
\(N_{\mathrm{slow}}\) is the number of slices assigned to the slower of the two adjacent ranks, 
\(T_t^{\mathrm{slow}}=\max(T_t^{-},T_t^{+})\) is the execution time of that slower rank, 
and \(\alpha\) limits the magnitude of the adjustment.

\subsection{RNN-Based Boundary Control}
\label{sec:rnn}

Rather than forecasting the full workload histogram explicitly, the RNN acts as a controller defined on partition boundaries. 
A single recurrent cell is shared by all interior boundaries, while each boundary $b$ maintains its own hidden state. 
At each rebalancing step, the hidden state is updated as
\begin{equation}
\mathbf{h}_{b,t}
=
f_{\mathrm{RNN}}
\left(
\mathbf{x}_{b,t},
\mathbf{h}_{b,t-1}
\right),
\end{equation}
where $\mathbf{x}_{b,t}$ denotes the feature vector observed at boundary $b$ at rebalancing step $t$, and $\mathbf{h}_{b,t}$ is the corresponding recurrent hidden state. 
Because the recurrent weights are shared and each input is defined locally at a boundary, the same trained controller can be applied to different numbers of GPU ranks.

The feature vector $\mathbf{x}_{b,t}$ summarizes the local state around boundary $b$. 
It includes timing imbalance between the two adjacent ranks, changes in local imbalance, global imbalance across ranks, the widths and remaining slack of the adjacent partitions, the previously applied boundary movement, the current phase of the cell-growth cycle, an indicator for recent growth events, and the reactive proposal $\Delta s_t^{\mathrm{r}}$. 
All features are expressed as dimensionless ratios and clamped to fixed ranges, avoiding the need to store normalization statistics for deployment.

The recurrent state is mapped to a scalar residual $r_{b,t}$, which corrects the reactive proposal in a bounded step space:
\begin{equation}
\Delta s_t
=
\Delta s_{\max}
\tanh\!\left(
\tanh^{-1}\!\left(
\frac{\Delta s_t^{\mathrm{r}}}{\Delta s_{\max}}
\right)
+ r_{b,t}
\right),
\end{equation}
where $\Delta s_{\max}$ bounds the maximum boundary movement during one rebalancing event. 
In implementation, the reactive proposal is clipped to the bounded range before applying the inverse hyperbolic tangent. 
This residual formulation allows the learned controller to refine the reactive rule while preserving a bounded adjustment magnitude.

The controller is implemented as a single-layer gated recurrent unit (GRU) with a hidden-state size of 24. 
A linear output layer maps the hidden state to the residual $r_{b,t}$. 
The output layer is initialized to zero, so before training the controller reproduces the reactive proposal. 
The complete controller contains approximately $2.8\times10^{3}$ trainable parameters.

\subsection{Residual Boundary Adjustment}

The corrected adjustments $\Delta s_t$ are applied jointly to all interior boundaries in a single sweep. 
Each boundary is moved toward the slower of its two adjacent ranks, and the movement is truncated if it would violate the minimum partition width required for ghost-region exchange. 
After the sweep, the updated boundaries define the subdomains used in the next simulation interval.

This formulation separates the direction and feasibility constraints from the learned residual. 
The reactive rule provides a baseline movement, the RNN modifies its magnitude through the residual correction, and the boundary-update sweep enforces admissible partition sizes before slice migration is performed.

\subsection{Training Strategy}
\label{sec:rnn_training}

The RNN controller is trained in a differentiable simulator of the load-balancing loop. 
The simulator reproduces the runtime rebalancing cadence and generates synthetic workload traces with randomized workload evolution, periodic growth events, transient workload bursts, and measurement jitter. 
During training, the controller observes noisy timing information, while the objective is evaluated using the underlying noise-free timing values. 
This discourages the controller from overreacting to measurement noise.

We use synthetic workload traces because the controller is intended to learn the structure of the boundary-adjustment policy rather than the detailed biological dynamics of a specific simulation run. 
The input features are dimensionless local quantities, such as timing imbalance, partition widths, growth-cycle information, and the reactive boundary proposal. 
This representation allows the controller to be trained across randomized workload scenarios and then evaluated in the multicellular simulation setting.

Each training episode samples a new scenario with randomized rank count, domain size, and minimum partition width. 
Evaluation uses scenario seeds that are not used during training, so temporally related observations from the same trajectory do not appear in both sets. 
Together with the dimensionless feature representation and shared per-boundary weights, this randomization supports generalization across problem geometries.

No fixed input window is used. 
Instead, the hidden state is carried across the simulation, and the network is trained by backpropagation through time over rollouts of 50 rebalancing events, corresponding to 5{,}000 simulation iterations at a rebalancing cadence of 100 iterations. 
Evaluation uses 100-event rollouts on unseen scenarios, with an additional 400-event rollout to check long-horizon stability.

The training loss averages the relative imbalance $(\max-\mathrm{mean})/\mathrm{mean}$ of the per-rank times and their normalized spread over the rollout. 
The spread term is weighted by 0.3, and both imbalance terms are normalized by a per-episode difficulty scale. 
An $L_1$ migration penalty of $5\times10^{-4}$ per moved slice is added to discourage unnecessary boundary movement.

Training uses the Adam optimizer with an initial learning rate of $3\times10^{-3}$, cosine annealing to 5\% of the initial value, gradient-norm clipping at 1.0, 6{,}000 optimization steps, and batches of 32 randomized episodes. 
The randomized scenarios use 2--16 ranks with varying domain sizes and minimum partition widths. 
A soft relaxation is used during training to keep the integer boundary assignment differentiable; its temperature is annealed from 1.5 to 0.4 so that the final policy is close to the hard integer assignment used during deployment. 
The exported controller weights are verified against the runtime implementation using recorded input--output parity tests.

The RNN is used only for dynamic rebalancing. 
The initial decomposition continues to use the computed workload scores described in Section~\ref{sec:initial_partition}.

\section{Experimental Evaluation}
\label{sec:evaluation}

We evaluate the proposed framework from four perspectives: GPU acceleration of element-level interactions, scalability across multiple GPUs, effectiveness of dynamic load balancing, and effectiveness of the learned RNN controller relative to the static and reactive load-balancing strategies, with additional comparisons to conventional time-series prediction baselines. We further present a representative multicellular growth use case to demonstrate the scientific capability enabled by the scalable framework.

\subsection{Experimental Setup}

All experiments use the epidermal subcellular element model described in Section~\ref{sec:framework} unless otherwise stated. 
The same biological and numerical parameters are used when comparing parallel configurations. 
All experiments were conducted on the Swan cluster at the Holland Computing Center (HCC), University of Nebraska--Lincoln. 
Swan is a Linux-based cluster with Intel Xeon Gold 6348 CPU nodes; each standard node contains two CPUs with 56 total CPU cores and 256~GB of system memory, connected through HDR100 InfiniBand. 
The GPU experiments used NVIDIA L40S GPU nodes, each equipped with four NVIDIA L40S GPUs with 48~GB of memory per GPU. 
Unless otherwise stated, each simulation rank was assigned to one GPU.

The multicellular simulation code was implemented using OpenCL for GPU computation and MPI for communication among ranks. 
MPI communication used OpenMPI through the Swan \texttt{openmpi/4.1} module. 
The OpenCL GPU runtime was provided through the NVIDIA GPU software stack available on the Swan L40S nodes, with \texttt{cuda/12.2} loaded for the experiments.
The RNN controller was implemented and trained using \texttt{pytorch/2.5.1}.

For the reactive load-balancing baseline, the imbalance threshold is set to $\tau=0.02$, and the boundary-adjustment coefficient is set to $\alpha=0.5$. 
The threshold controls when repartitioning is triggered, while $\alpha$ limits the magnitude of each boundary adjustment. 
All multi-GPU experiments partition the spatial domain along the $x$-axis, which is selected to better match the tissue geometry used in our experiments.

The RNN controller architecture and training procedure are described in Sections~\ref{sec:rnn} and~\ref{sec:rnn_training}.

\begin{figure}[t]
    \centering
    \includegraphics[width=0.80\columnwidth]{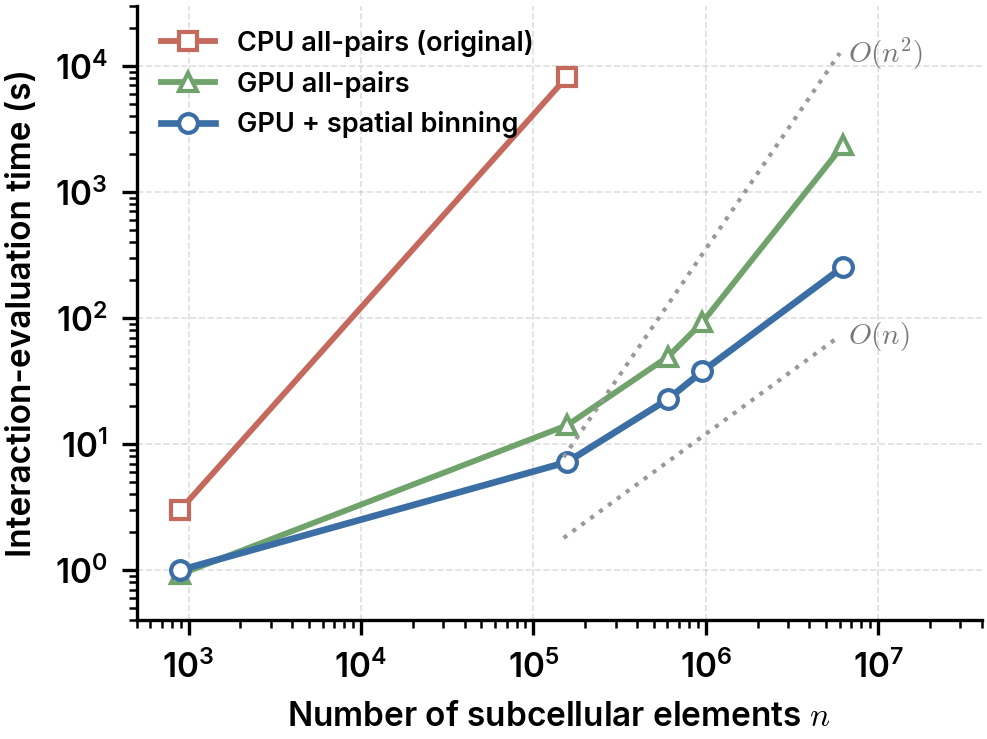}
    \caption{Single-GPU performance of element-level interaction evaluation measured over \(1{,}000\) simulation iterations on a log--log scale as the number of subcellular elements increases.
    The figure compares the original serial CPU all-pairs implementation, the GPU all-pairs implementation, and the GPU implementation with spatial binning. 
    Reference slopes indicate the expected $O(n^2)$ and $O(n)$ scaling trends.}
    \label{fig:single_gpu}
\end{figure}

\subsection{Single-GPU Acceleration}

We first evaluate GPU acceleration and spatial binning independently of multi-GPU execution. 
Figure~\ref{fig:single_gpu} reports the element-level interaction-evaluation time as the number of subcellular elements increases.

We compare three implementations of the element-level interaction evaluation: the original serial CPU all-pairs implementation, a GPU-parallel all-pairs implementation, and a GPU implementation with spatial binning. 
The implementations are evaluated using the same set of input configurations, ranging from 60 cells ($891$ elements) to $435{,}661$ cells ($6.3\times10^{6}$ elements). 
The two GPU implementations are run on a single NVIDIA L40S GPU, while the serial implementation is run on the host Intel Xeon Gold 6348 CPU.

The serial CPU implementation becomes impractical well before the largest problem sizes. 
At $1.6\times10^{5}$ elements, it requires roughly 2 hours, whereas the GPU all-pairs implementation completes in $14.1$~s and the binned GPU implementation completes in $7.2$~s, yielding a speedup of roughly three orders of magnitude over the CPU baseline. 
Larger configurations were not attempted on the CPU.

Moving the all-pairs evaluation to the GPU removes the serial bottleneck but retains the quadratic work complexity, and the measured runtime grows superlinearly with element count. 
Spatial binning restricts force evaluation to a bounded local neighborhood, producing near-linear scaling under bounded local density. 
For example, increasing the element count by approximately $40\times$, from $1.6\times10^{5}$ to $6.3\times10^{6}$, increases the binned GPU runtime by only $35\times$. 
At the largest configuration, spatial binning completes in $255$~s, compared with $2{,}355$~s for GPU all-pairs, giving a $9.2\times$ speedup. 
Only for the smallest input, with fewer than $10^{3}$ elements, does the binning overhead offset its benefit, making the two GPU implementations comparable.

\begin{figure}[t]
    \centering
    \includegraphics[width=0.80\columnwidth]{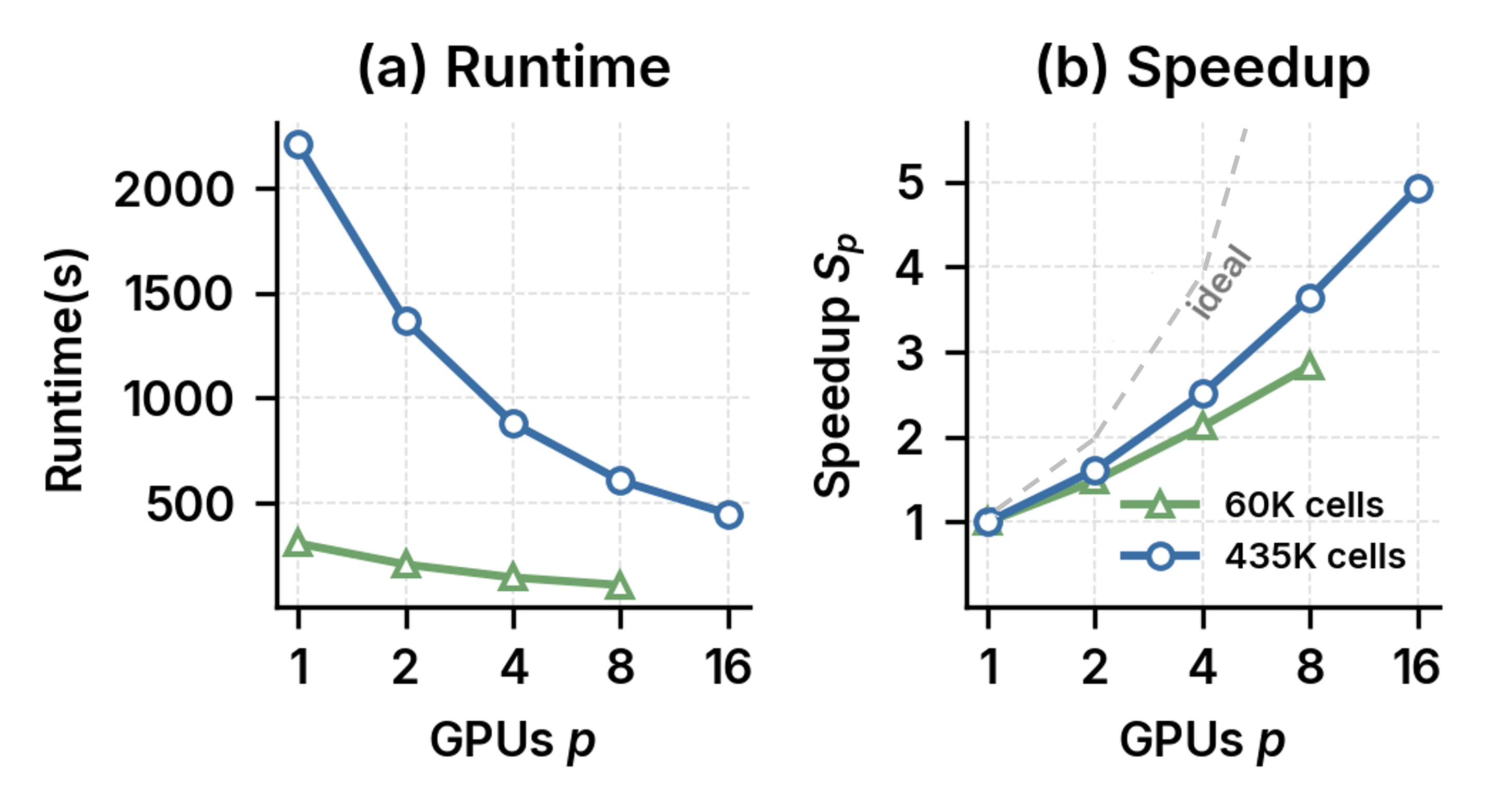}
    \caption{Multi-GPU strong-scaling performance of GPU computation time for two multicellular simulation sizes. 
    (a) GPU computation time as the number of GPUs increases. 
    (b) Corresponding speedup, with the dashed line indicating ideal linear scaling.}
    \label{fig:scaling}
\end{figure}

\subsection{Multi-GPU Scalability}

We next evaluate the strong-scaling performance of the GPU computation component of the multi-GPU simulation framework on NVIDIA L40S GPUs. 
Each simulation rank is assigned to one GPU. 
Figure~\ref{fig:scaling} reports GPU computation time and the corresponding computation-only speedup for two fixed multicellular simulation sizes as the number of GPUs increases. 
The reported time excludes startup, load-balancing, communication, and I/O costs, and therefore isolates the scaling behavior of the core GPU computation.

For $p$ GPUs, the computation-only speedup is
\begin{equation}
S_p = \frac{T_1}{T_p},
\end{equation}
where $T_1$ and $T_p$ are the GPU computation times using one and $p$ GPUs, respectively. 
The corresponding parallel efficiency is
\begin{equation}
E_p = \frac{S_p}{p}.
\end{equation}

Figure~\ref{fig:scaling} shows that GPU computation time decreases as additional GPUs are used, demonstrating that the core interaction computation benefits from distributed execution. 
The speedup remains sublinear for both simulation sizes, as expected in a strong-scaling experiment with fixed problem sizes. 
For the 60K-cell case, the workload becomes too small to fully utilize each GPU as the number of GPUs increases, and scaling begins to saturate by 8 GPUs. 
The larger 435K-cell case provides more computation per rank and therefore achieves better GPU utilization at the same GPU count. 
For example, at 8 GPUs, the 435K-cell case reaches approximately \(45\%\) computation-only parallel efficiency, compared with \(36\%\) for the 60K-cell case.

The 435K-cell case continues to benefit from additional GPUs up to 16 GPUs, but its efficiency drops to \(31\%\) as the fixed workload is divided across more devices. 
This indicates that, even for the larger case, the per-GPU workload at 16 GPUs is not sufficient to fully saturate all GPUs. 
These results show that the core GPU computation benefits from multi-GPU execution, and also suggest that larger production-scale simulations are better suited to fully utilize additional GPUs. 
Full simulation performance, including dynamic repartitioning, is evaluated separately in the following experiments.

\subsection{Dynamic Load-Balancing Performance}

\begin{figure}[t]
    \centering
    \includegraphics[width=0.8\linewidth]{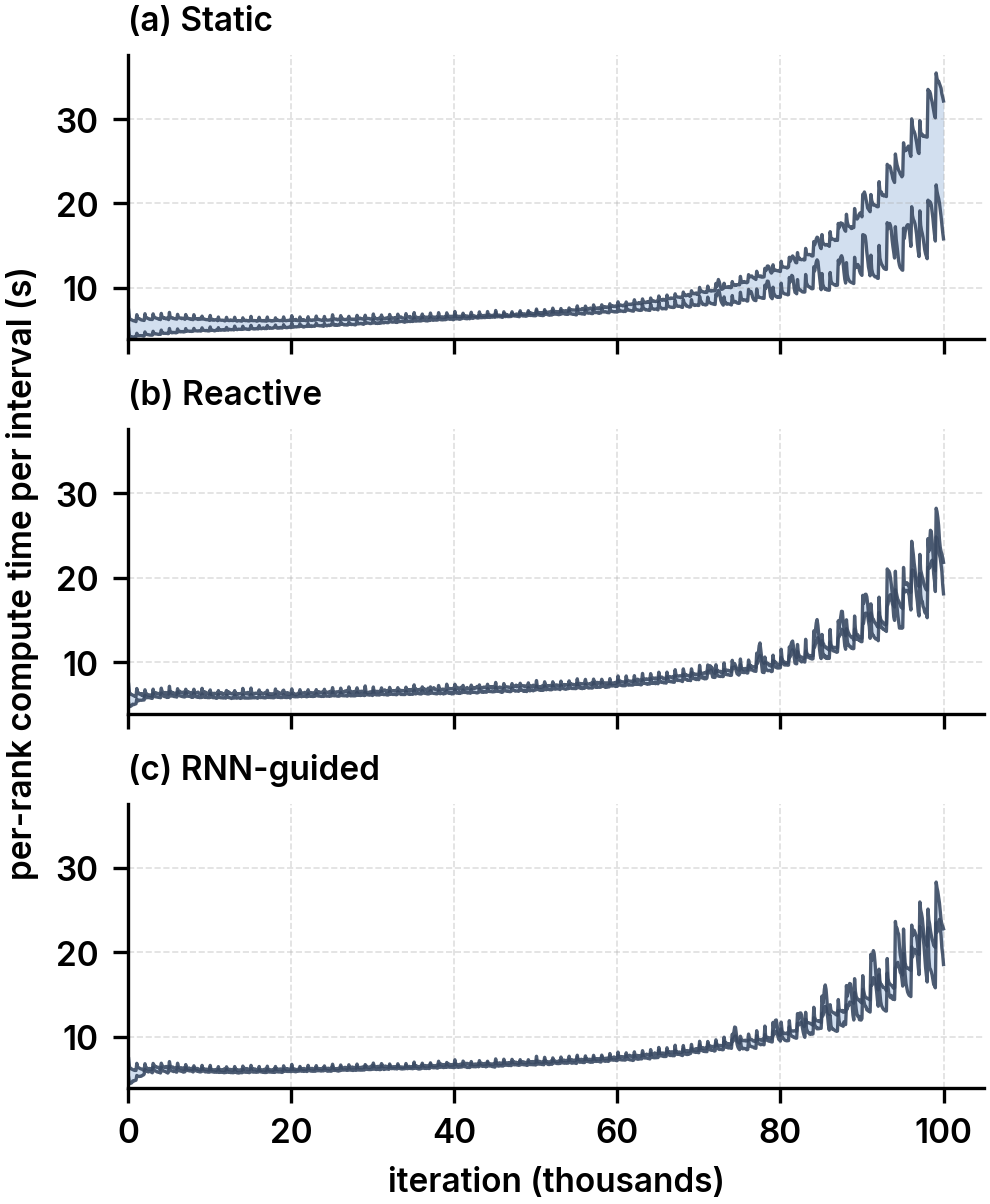}
    \caption{Evolution of per-rank computation time during the simulation for static partitioning, reactive load balancing, and RNN-guided load balancing over a \(100{,}000\)-iteration simulation. 
    Each curve represents one GPU rank, and each point reports the rank's computation time over one simulation interval. 
    The x-axis reports simulation iterations in units of \(10^3\).}
    \label{fig:rank_balance}
\end{figure}

\begin{figure}[t]
    \centering
    \includegraphics[width=0.98\linewidth]{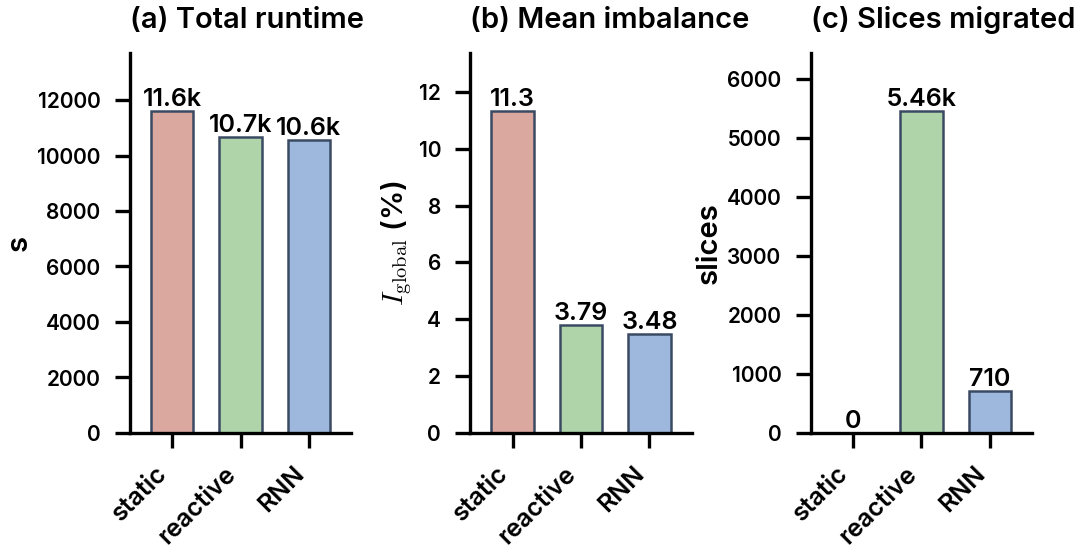}
    \caption{
    End-to-end performance of static partitioning, reactive load balancing, and RNN-guided load balancing over a \(100{,}000\)-iteration simulation.    
    The figure compares total runtime, mean global imbalance, and slice migration cost. 
    Dynamic load balancing reduces timing imbalance, while slice migration measures the amount of data movement introduced by repartitioning.}
    \label{fig:load_balance}
\end{figure}

We compare three multi-GPU configurations: static partitioning, reactive load balancing, and the proposed RNN-guided method. 
The goal is to determine whether dynamic repartitioning reduces the delay caused by the slowest GPU rank and whether improved load balance translates into end-to-end performance gains. 
Figure~\ref{fig:rank_balance} shows the evolution of per-rank computation times during the simulation, while Fig.~\ref{fig:load_balance} summarizes total runtime, mean global imbalance, and slice migration cost.

A global timing imbalance is measured as
\begin{equation}
\label{eq:imbalance}
I_{\mathrm{global}}
=
\frac{T_{\max}-T_{\min}}
{T_{\mathrm{mean}}},
\end{equation}
where $T_{\max}$, $T_{\min}$, and $T_{\mathrm{mean}}$ are the maximum, minimum, and mean computation times across GPU ranks for a simulation interval. 
The mean global imbalance reported in Fig.~\ref{fig:load_balance} is computed by averaging $I_{\mathrm{global}}$ over the measured intervals.

Figure~\ref{fig:rank_balance} shows that static partitioning produces persistent timing differences among GPU ranks, causing execution to be limited by the slowest rank. 
Reactive load balancing reduces this spread after imbalance is observed, while the RNN-guided method further tightens the rank computation-time curves and slightly lowers per-rank computation times. 
This indicates that the learned residual corrections improve workload balance beyond the purely reactive rule.

Figure~\ref{fig:load_balance} further summarizes these effects. 
Both dynamic methods reduce the mean global imbalance compared with static partitioning, and the RNN-guided method achieves the lowest imbalance among the three configurations. 
The RNN-guided method also reduces total runtime relative to static partitioning and achieves runtime comparable to, and slightly lower than, the reactive baseline in this experiment.

The main difference between the two dynamic methods appears in the amount of slice migration. 
Reactive load balancing reduces imbalance by frequently shifting boundaries, which leads to a larger number of migrated slices. 
In contrast, the RNN-guided method maintains tighter per-rank timing balance while migrating far fewer slices. 
This suggests that the learned controller performs repartitioning more selectively, avoiding unnecessary boundary movement while preserving balanced GPU workloads. 
Overall, these results show that the RNN-guided controller improves the balance--migration tradeoff by maintaining balanced GPU workloads with comparable end-to-end runtime to the reactive baseline while substantially reducing the amount of data movement required during repartitioning.

\subsection{RNN Controller Evaluation}
\label{sec:rnn_eval}

Because the RNN controller outputs boundary adjustments rather than an explicit workload forecast, we evaluate it by the quality of the resulting load balance. 
The controller is compared with static partitioning, the reactive baseline, and two conventional forecast-then-repartition policies based on persistence and Holt-based prediction~\cite{alkharboush2013holt,degrande2017timeseries}. 
All methods are evaluated on the same held-out scenarios using hard integer partition boundaries.

For a rollout with \(K\) rebalancing steps, we measure the time-averaged global imbalance as
\begin{equation}
\bar{I}_{\mathrm{global}}
=
\frac{1}{K}
\sum_{t=1}^{K}
I_{\mathrm{global},t},
\end{equation}
where \(I_{\mathrm{global},t}\) is the global timing imbalance at rebalancing step \(t\), defined in Eq.~\ref{eq:imbalance}. 
We also measure the migration cost as the total number of slices moved during the rollout:
\begin{equation}
M
=
\sum_{t=1}^{K}
\sum_{b \in \mathcal{B}}
|\Delta s_{b,t}|,
\end{equation}
where \(\mathcal{B}\) is the set of interior partition boundaries and \(\Delta s_{b,t}\) is the number of slices moved at boundary \(b\) during rebalancing step \(t\). 
Thus, \(M\) measures total slice migration over the full rollout, rather than mean migration per step.

\begin{figure}[t]
    \centering
    \includegraphics[width=0.98\columnwidth]{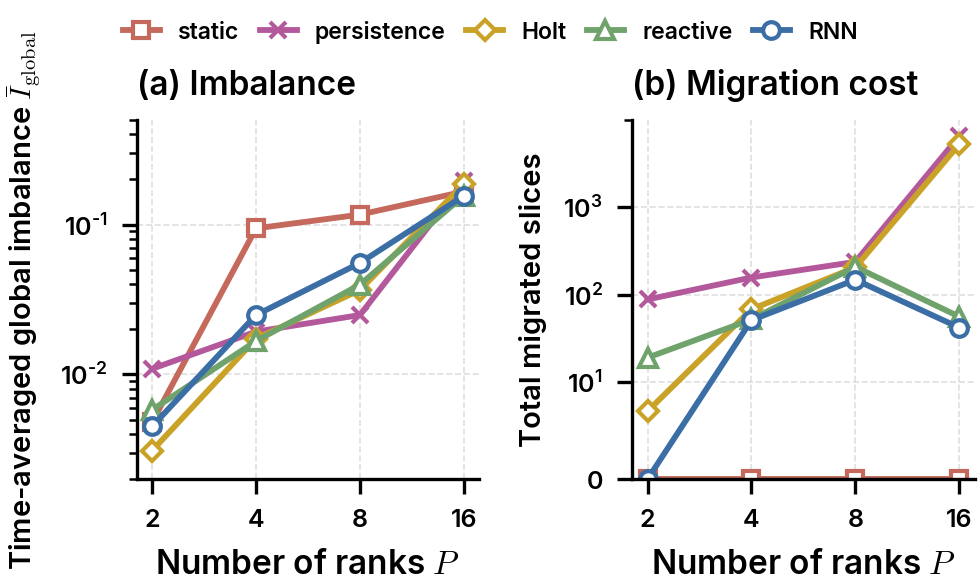}
    \caption{Controller-level evaluation on held-out scenarios. 
    (a) Time-averaged global imbalance for static partitioning, persistence-based forecast-then-repartition, Holt-based forecast-then-repartition, the reactive baseline, and the RNN controller. 
    (b) Migration cost measured as the total number of slices migrated over the rollout. 
    Each point corresponds to one held-out rollout on the 435K-cell initial condition, with imbalance averaged over the measured intervals and migration accumulated over the rollout. 
    The vertical axis is logarithmic in both panels, with a linear region below \(10\) in (b) so that static partitioning, which migrates no slices, remains visible. 
    Lower values are better for both metrics.}
    \label{fig:rnn_prediction}
\end{figure}

Figure~\ref{fig:rnn_prediction} compares the policies across different rank counts. 
Static partitioning incurs no migration cost, but it cannot adapt to changes in workload distribution. 
The forecast-then-repartition methods and the reactive baseline reduce imbalance by updating partition boundaries, but they can require substantially more slice migration. 
The RNN controller is evaluated under the same hard-boundary setting to determine whether its learned residual corrections improve the balance--migration tradeoff.

At \(P=2\), the static partition is already balanced to within \(0.5\%\), leaving little imbalance for any controller to correct. 
In this case, additional boundary movement can be counterproductive, as shown by the persistence policy, which migrates \(88\) slices and increases the time-averaged imbalance. 
At \(P=4\) and \(P=8\), adaptive methods reduce imbalance relative to static partitioning, although the best imbalance is not always achieved by the RNN controller. 
At \(P=16\), the forecast-then-repartition policies become less effective because the slices are thin and frequent boundary movement introduces excessive migration. 
Persistence and Holt migrate \(6481\) and \(5243\) slices, respectively, while producing higher imbalance than static partitioning. 
In contrast, the reactive and RNN controllers keep imbalance at or below the static level using fewer than \(60\) slice migrations.

Across all rank counts, the RNN controller migrates the fewest slices among the adaptive methods while keeping time-averaged imbalance below the static baseline. 
Although it does not always achieve the lowest imbalance at moderate rank counts, it provides the most conservative migration schedule. 
These results indicate that the learned residual corrections trade a small amount of imbalance reduction for a substantially lower migration cost.

In addition to the standard held-out rollouts, we evaluate a longer rollout to check that the recurrent hidden state remains stable beyond the training horizon. 
Recorded input--output parity tests are also used to verify that the exported controller weights reproduce the training-time model in the runtime implementation.

Controller-level gains alone do not determine end-to-end usefulness. 
The results in Figs.~\ref{fig:rank_balance} and~\ref{fig:load_balance} therefore examine whether the imbalance reduction observed in this isolated controller evaluation translates into improved multi-GPU simulation behavior.

\subsection{Scientific Use Case: Embryonic Epidermal Development}

\begin{figure}[t]
    \centering
    \includegraphics[width=0.95\linewidth]{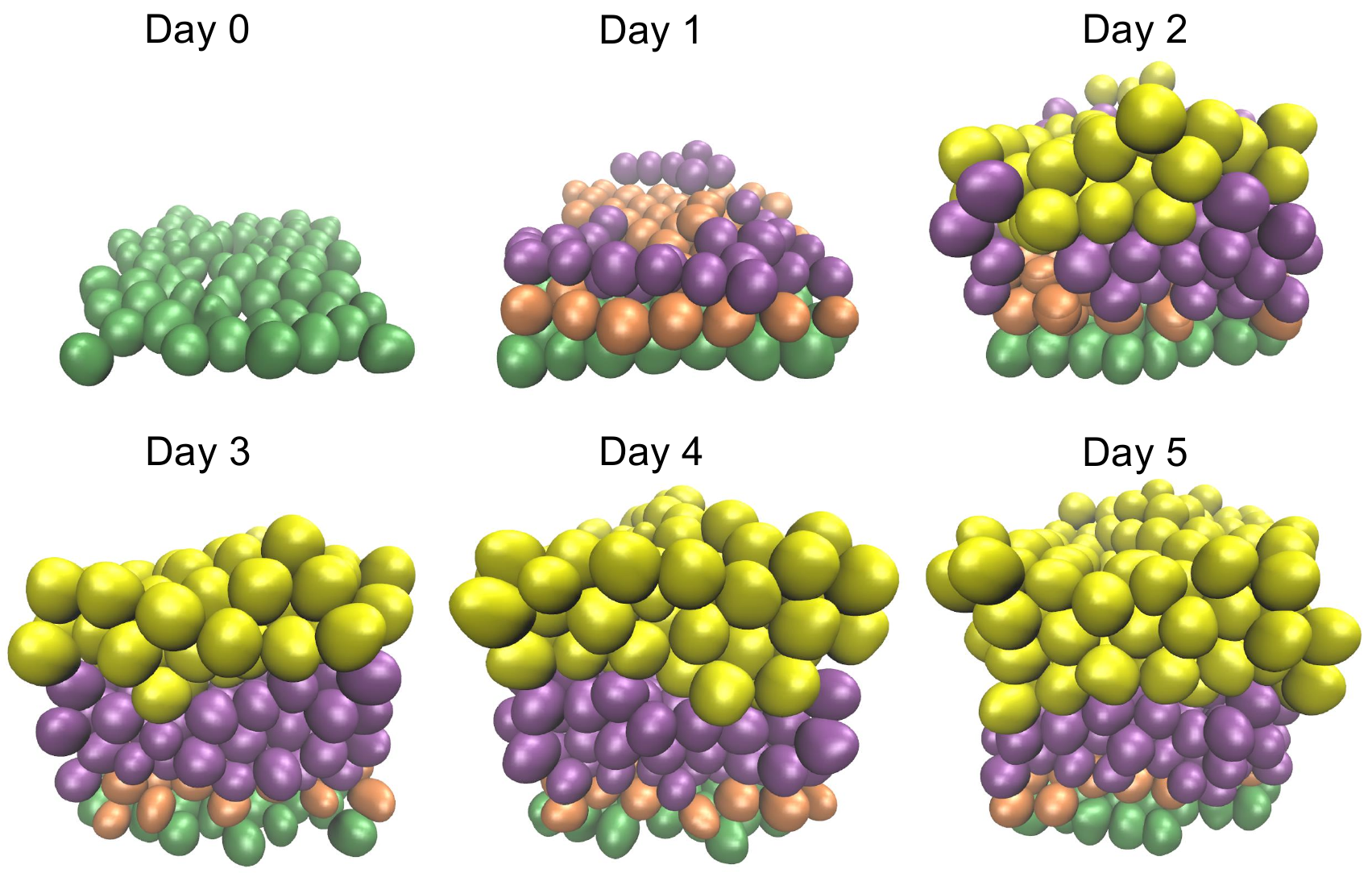}
    \caption{Representative epidermal development simulation at multiple time points. 
    Cells are rendered as colored ellipsoids, with different colors indicating different cell types or states. 
    The snapshots show the evolving cell population over a 5-day developmental period, illustrating progressive proliferation, differentiation, and tissue reorganization. 
    The resulting spatially and temporally nonuniform workload motivates the proposed multi-GPU framework and dynamic load-balancing strategy.}
    \label{fig:epidermal_visualization}
\end{figure}

We use embryonic epidermal development as a representative scientific use case for scalable multicellular growth simulation.
During embryogenesis, epidermal cells proliferate, migrate, and differentiate to form distinct tissue layers through stratification, producing a spatially heterogeneous cell population whose composition and local density evolve over developmental time~\cite{du2018multiscale}.
Figure~\ref{fig:epidermal_visualization} shows representative snapshots of the simulation from Day~0 to Day~5, illustrating the emergence of a multilayered tissue structure as the cell population grows and differentiates.
This example is not intended as a separate biological validation study. 
Instead, it highlights the type of spatially evolving multicellular workload targeted by the proposed multi-GPU framework.

\section{Conclusion}
\label{sec:conclusion}

This work presents a scalable multi-GPU framework for three-dimensional multicellular growth simulation based on a subcellular element model. 
The framework combines GPU acceleration, spatial binning, domain decomposition, and workload-aware initial partitioning to address the high cost of element-level interactions and the spatially nonuniform workload of a growing tissue.

A key challenge is that cell movement, growth, and division continuously alter the workload distribution, causing an initially balanced decomposition to become inefficient over time. 
To address this problem, we introduce an RNN-guided load-balancing controller that learns residual corrections to a reactive boundary-adjustment rule from recent workload and partition-state history. 
Rather than simply reacting to observed imbalance with frequent boundary movement, the learned controller uses temporal information to make more selective repartitioning decisions.

Quantitatively, GPU acceleration with spatial binning reduces the element-interaction time by roughly three orders of magnitude relative to a serial CPU baseline, and binning provides a further \(9.2\times\) speedup over a GPU all-pairs implementation at \(6.3\times10^{6}\) elements. 
In end-to-end 435K-cell simulations on 8 GPUs, RNN-guided load balancing reduces the mean global imbalance from \(11.3\%\) under static partitioning to \(3.5\%\), compared with \(3.8\%\) for the reactive rule, and lowers total runtime by \(9.0\%\) relative to static partitioning. 
Compared with the reactive baseline, the RNN-guided controller achieves slightly better load balance with comparable runtime while migrating \(7.7\times\) fewer slices, demonstrating a substantially improved balance--migration tradeoff.

The embryonic epidermal development use case further demonstrates how the scalable framework can support dynamically evolving tissue simulations in which proliferation, differentiation, and tissue reorganization create spatially and temporally nonuniform computational workloads. 
Overall, the proposed approach provides a practical path toward efficient simulation of dynamically evolving multicellular systems on modern multi-GPU platforms, while showing that history-aware load balancing can reduce unnecessary repartitioning without sacrificing workload balance.

In future work, we plan to extend the evaluation to larger production-scale simulations, including million-cell-level tissue models. 
The current strong-scaling experiments use fixed problem sizes, and at high GPU counts the per-GPU workload may become too small to fully utilize all devices. 
Larger simulations with more cells and subcellular elements are expected to provide more computation per GPU, better expose the benefits of distributed execution, and further test the scalability of the proposed load-balancing strategy.

\bibliographystyle{IEEEtran}
\bibliography{references}

\end{document}